\documentclass[prl,aps,twocolumn]{revtex4}
\usepackage[dvips]{graphicx}
\begin{document}

\title{Noise suppression in reproduction of single flux quantum pulses}
\author{Anna V. Gordeeva and Andrey L. Pankratov}
\email[]{alp@ipm.sci-nnov.ru}
\affiliation{Institute for Physics of Microstructures of RAS,
Nizhny Novgorod, 603950, Russia.}

\begin{abstract}
We present the analysis of the mean switching time and its standard deviation of an
overdamped Josephson junction, driven by a direct current and a single flux quantum
(SFQ) pulse. The performed analysis allows to find the optimal value of the bias
current of the clock generator, responsible for the shape of SFQ pulse, which
minimizes noise-induced switching errors.
\end{abstract}

\maketitle

The investigation of the fluctuational characteristics of Josephson junctions (JJ) is
very important due to their applications as logic devices \cite{rsfq},\cite{rl}.
Existence of fluctuations in Josephson junctions leads, for example, to limiting
lifetimes of the information unit, recorded in Josephson memory cell, to random
switching of logic gates and to spread of arrival time of signals in Josephson
transmission lines (see, for example, \cite{rl} and \cite{lik}). In the present paper
we consider the dynamics of a short overdamped JJ driven by a direct current and SFQ
pulse. An SFQ pulse is supposed to be born by another JJ named a clock generator. We
present the analysis of the mean switching time (MST) and its standard deviation (SD)
of the overdamped JJ versus bias current of the clock generator, that is responsible
for the shape of SFQ pulse. The performed investigation allows to find the optimal
value of the bias current of the clock generator, which minimizes switching errors
induced by noise. In addition we test the limits of applicability of the formula
for the SD recently derived by Semenov and Inamdar \cite{sem}.

It is well-known, that in the frame of Resistively-Shunted-Junction model \cite{lik}
a point Josephson junction in the limit of a small capacitance
(high damping), driven by current $I$ with fluctuations taken into
account is well described by the Langevin equation:
\begin{equation}
\label{s1}\omega _c^{-1}{\frac{d\varphi (t)}{dt}}=-{\frac{du(\varphi )}{d\varphi }}-i_F(t),
\end{equation}
here $u(\varphi )=1-\cos \varphi -i(t)\varphi$ is the dimensionless potential profile,
$\varphi$ is the difference in the phases of order parameter on opposite sides of the
junction, $\displaystyle{i(t)=i_0+f(t)=I/{I_c}}$, $I_c$ - critical current of JJ,
$\displaystyle{i_F(t)={I_F }/{I_c}}$, $I_F$ is the random component of the current,
$\displaystyle \omega_c={{2eR_NI_c}/\hbar }$ is the characteristic frequency of the
junction. In the case when only
thermal fluctuations are taken into account, the random current may be represented by
white Gaussian noise: $\langle i_F(t)\rangle=0,\quad \displaystyle{\langle
i_F(t)i_F(t+\tau)\rangle ={\frac{2\gamma }{\omega _c}}\delta (\tau )}$, where
$\gamma=~2ekT/\hbar I_c$ is the dimensionless intensity of fluctuations, $e$ is the
electron charge, $k$ is the Boltzmann constant, $T$ is the temperature, and  $\hbar$
is the Planck constant.

Initially, a current going across the junction is smaller than the critical one,
$i_0<1$. A single flux quantum pulse arriving from a clock generator (or simply from
another JJ) switches the junction to the resistive state, that leads to generation of
new SFQ pulse. This pulse will be generated not immediately, but at the later time
which is called the switching time. Since due to noise the moment of pulse generation
is a random quantity, let us investigate its mean and standard deviation. As a clock
pulse we choose a voltage pulse, shape of which can be obtained by solution of
equation (\ref{s1}) without any fluctuations for $i>1$:
$f(t)=A\left(\frac{a\omega^2}{a^2-\cos(\omega t+\psi)+\omega\sin(\omega
t+\psi)}+1-a\right)$, where $a$ is the current going across the junction utilized as a
clock generator, $\omega=\sqrt{a^2-1}$ is the oscillation frequency,
$\psi=\pi+\arctan(-\omega)$, $A$ is the signal magnitude. In the inset of Fig. 1 the
form of the current pulse $i(t)=i_0+f(t)$ is presented for the case of the equal
maximal value of current $i(t)=2.9$.
\begin{figure}[h]
\begin{picture}(220,150)(10,10)
\centering\includegraphics[width=8cm,height=6cm]{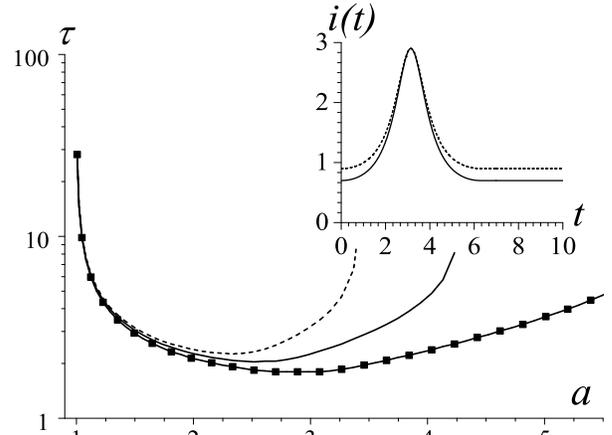}
\end{picture}
\parbox{230bp}{\caption{
The MST versus current $a$ for $\gamma=0$ and different bias currents:
$i_0=0.9$, $A=1$ - line with squares; $i_0=0.8$, $A=1.05$ - solid line;
$i_0=0.7$, $A=1.1$ - dashed line.
Inset: the form of current pulse $i(t)$ for $A=1.05$, $i_0=0.8$ - solid line,
$A=1$, $i_0=0.9$ - dashed line.}
\label{Fig1}}
\end{figure}

By the definition \cite{ACP},\cite{PRL} the first and the second moments of switching
time take the following form: \\
\begin{equation}
\tau=\left< t \right >=\int_0^\infty t w(t)dt,\,\,
\left< t^2 \right >=\int_0^\infty t^2 w(t)dt,
\label{tau}
\end{equation}
where $w(t)= \partial P(t)/\partial t$, $P(t)$ is the probability to find $\varphi$
whithin the interval ($-\pi$, $\pi$) and the standard deviation is, as usual,
$\sigma=\sqrt{\left< t^2 \right > - \left< t \right >^2}$. In the following $\tau$ and
$\sigma$ will be determined both via direct computer simulations of Eq. (\ref{s1}) and
by numerical solution of the corresponding Fokker-Planck equation using the
Crank-Nicolson scheme. In all figures both $\tau$ and $\sigma$ are
normalized to $1/\omega_c$. Note, that for presently used
technological processes $1/\omega_c$ is of the order of $1$ $ps$ (see \cite{rl},\cite{sem}).

In Fig. 1 the MST is presented versus current $a$ in the case of zero noise intensity
for three different values of bias current $i_0$, but constant total current. MST has a
minimum as a function of current $a$, which for larger bias current $i_0$ becomes more
broad and deep: minimal MST increases of about 1.2 times for bias current $i_0=0.7$ in
comparison with $i_0=0.9$.
\begin{figure}[h]
\begin{picture}(220,150)(10,10)
\centering\includegraphics[width=8cm,height=6cm]{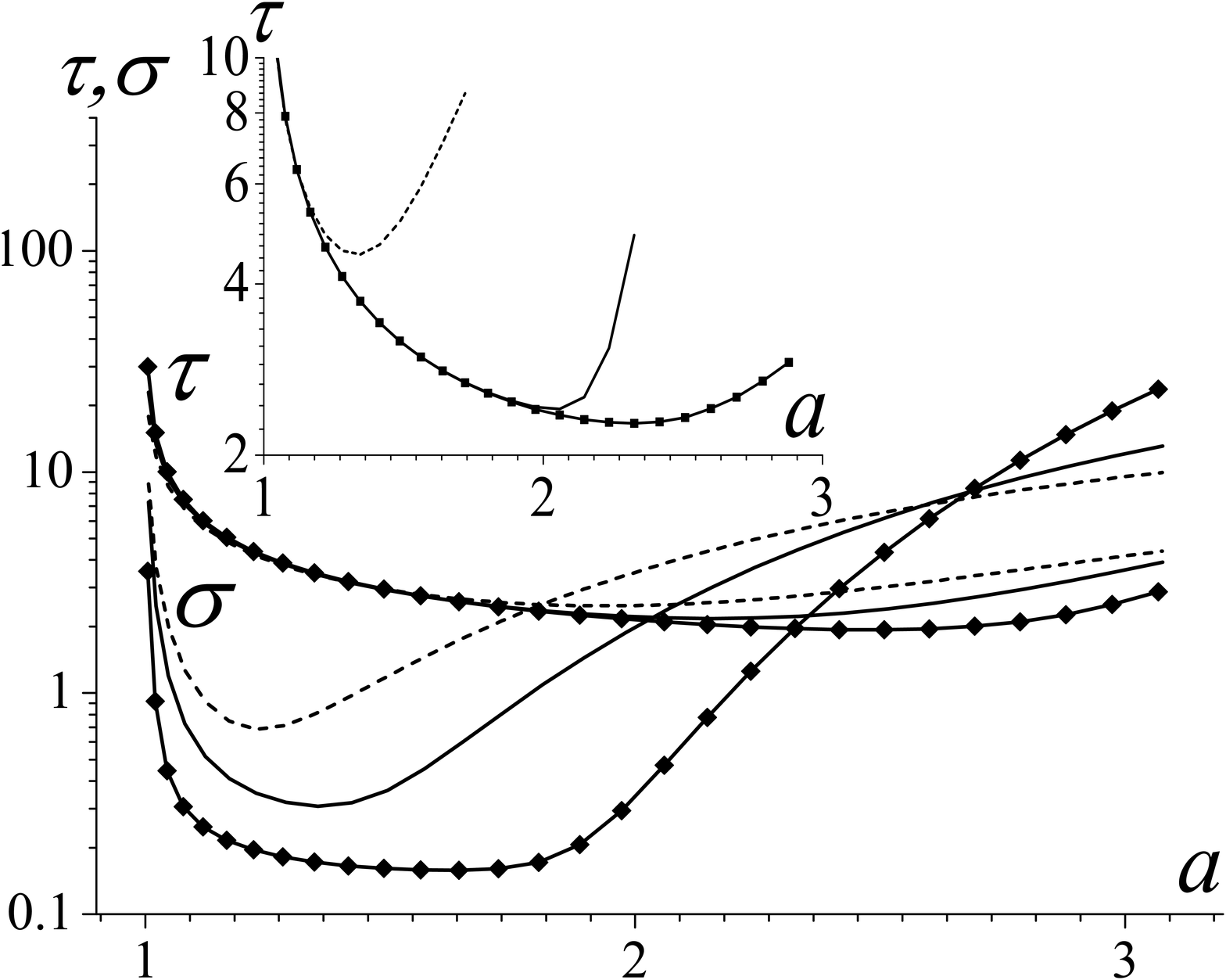}
\end{picture}
\parbox{230bp}{\caption{
The MST and SD versus current $a$ for different noise intensities: $\gamma=0.1$ -
dashed line, $\gamma=0.05$ - solid line, $\gamma=0.01$ - line with diamonds,
$i_0=0.9$, $A=1$.
Inset: $i_0=0.7$, $A=1.1$, $\gamma=0.001$ -
line with squares, $\gamma=0.01$ - solid line, $\gamma=0.1$ - dashed line.
}
\label{Fig2}}
\end{figure}

In Fig. 2 MST and SD versus current $a$ are presented for different noise intensities
for $i_0=0.9$, $A=1$. As it is seen, these characteristics have minima as functions of
$a$, and the minimum of $\tau$ is shifted to larger values of current $a$, and is more
broad and flat than the minimum of $\sigma$. One can see that for noise intensity
$\gamma=0.05$ the most optimal range $a$ is located around 1.3-1.4, because here the
minimum of $\sigma$ is reached, and MST is near the minimum and does not almost depend
on the noise intensity (curves $\tau$ for $\gamma=0.1-0.01$ around $a=1.4$ actually
coincide). Increasing of MST minimum with increasing of noise intensity is due to the
noise delayed decay effect that has been studied in \cite{PRL}, \cite{MP} in
connection with Josephson junctions. More clearly this effect appears, if one takes
smaller bias current $i_0=0.7$ and changes noise intensity in broader interval
($\gamma=0.1$, 0.01 and 0.001). As it is seen in the inset of Fig. 2 the location of
MST minimum and also its value significantly depends on noise intensity: minimal MST
for $\gamma=0.1$ is about 2 times larger than the minimal MST for $\gamma=0.001$.
Therefore, high-$T_c$ devices must be redesigned with account of this effect to get
maximal performance.

In Fig. 3 MST and SD are presented versus generator current $a$ for different values of
bias current $i_0$ and signal magnitude $A$, but constant total current $i_0+2A=2.9$. It is
seen, that for larger bias current $i_0$ switching occurs faster and is less random:
minimum of $\sigma$ is wider for $i_0 = 0.9$ and $A=1$ than for $i_0 = 0.7$ and $A=1.1$. It is
necessary to point, that minimal SD value does not depend on bias current, and is
determined by the noise intensity only (see also Fig. 4).
\begin{figure}[h]
\begin{picture}(220,150)(10,10)
\centering\includegraphics[width=8cm,height=6cm]{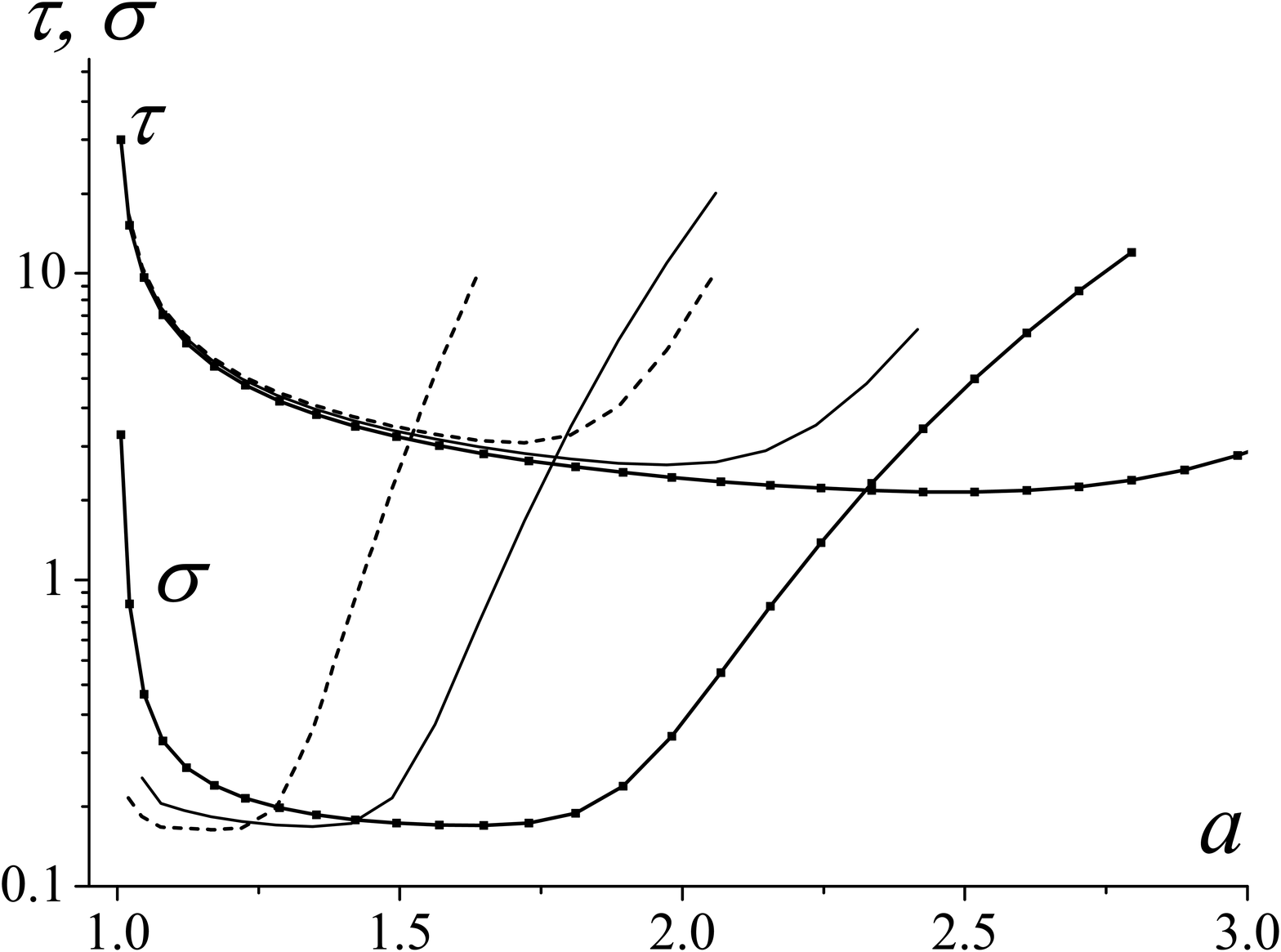}
\end{picture}
\parbox{230bp}{\caption{The MST and SD versus generator current $a$ ($\gamma=0.02$):
$i_0=0.9$, $A=1$ - line with squares; $i_0=0.8$, $A=1.05$ - solid line;
$i_0=0.7$, $A=1.1$ - dashed line. }
\label{Fig3}}
\end{figure}

In Fig. 4 the SD versus generator current $a$ is presented for $i_0=0.9$ and noise
intensities $\gamma=0.1;0.01;0.001$. One can see, that for
$\gamma\le 0.01$ minimal SD value scales as the square root of noise intensity as for
the case of a step-wise signal \cite{rl},\cite{PRL}. The asymptotic expression of SD,
derived in \cite{PRL} for the case $\gamma\ll 1$ (that is somewhat more exact than one,
derived in \cite{rl}) has the form:
\begin{eqnarray}\label{sg}
\sigma(\varphi_0)=\frac{1}{\omega_c}\sqrt{2\gamma
\left[F(\varphi_0)+f_3(\varphi_0)\right]+\dots },\\
\begin{array}{ccc}
F(\varphi_0)&=&f_1(\varphi_2)f_2(\varphi_2)-2f_1(\varphi_2)f_2(\varphi_0)+ \nonumber \\
& &f_1(\varphi_0)f_2(\varphi_0)+\frac{f_1(\varphi_2)-f_1(\varphi_0)}{(i-\sin(\varphi_0))^2},
\nonumber \\
f_1(x)&=&\frac 2{\sqrt{i^2-1}}
\arctan\left( \frac{i\tan(x/2)-1}{\sqrt{i^2-1}}\right), \\
\nonumber
f_2(x)&=&1/({2(i-\sin{x})^2}), \\
f_3(\varphi_0)&=& \int_{\varphi_0}^{\varphi_2}
\left[\frac{\cos(x)f_1(x)}{(\sin(x)-i)^3}-\frac{3}{2(\sin(x)-i)^3}
\right]dx.
\end{array}
\end{eqnarray}
The asymptotic values of SD, given by (\ref{sg}), are presented in Fig. 4 as dashed
straight lines: for $\gamma\le 0.01$ they are close to the minimum of $\sigma$, the
disagreement does not exceed $20 \%$. In paper \cite{Fil} it has been demonstrated,
that the jitter can be minimized by the use of the sharp pulses. As it follows from
the performed analysis (see Fig. 4), even for soft SFQ pulses the jitter can be
suppressed down to the level of sharp driving case if the parameters are properly
optimized.
\begin{figure}[h]
\begin{picture}(220,150)(10,10)
\centering\includegraphics[width=8cm,height=6cm]{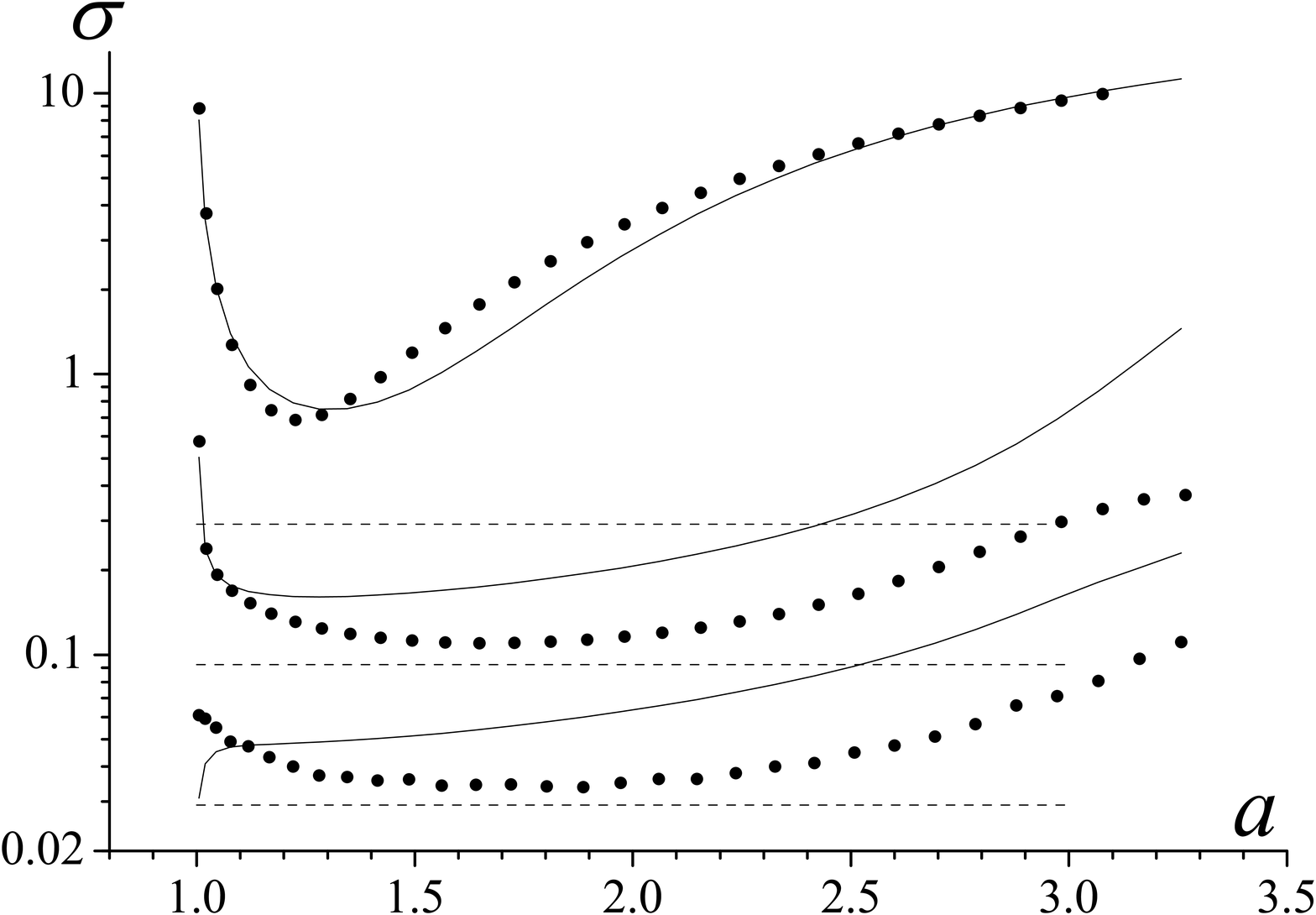}
\end{picture}
\parbox{230bp}{\caption{The SD versus current $a$ for $i_0=0.9$, $A=1$ and different
values of noise intensity, from top to bottom $\gamma=0.1;0.01;0.001$: circles -
computer simulations, solid curves - formula (\ref{sem}), dashed straight
lines - formula (\ref{sg}).}
\label{Fig4}}
\end{figure}

Fig. 5 shows the results of computer simulations for values of bias current and noise
intensity that are typical for real RSFQ circuits ($A=1$, $\gamma=0.001$,
$i_0=0.5;0.7;0.9$). As it is seen, for bias current equals 0.5 and 0.7  the minimum
of standard deviation is not observed in the considered range of parameters. This
could be a reason why this effect had not been observed before.
\begin{figure}[h]
\begin{picture}(220,150)(10,10)
\centering\includegraphics[width=8cm,height=6cm]{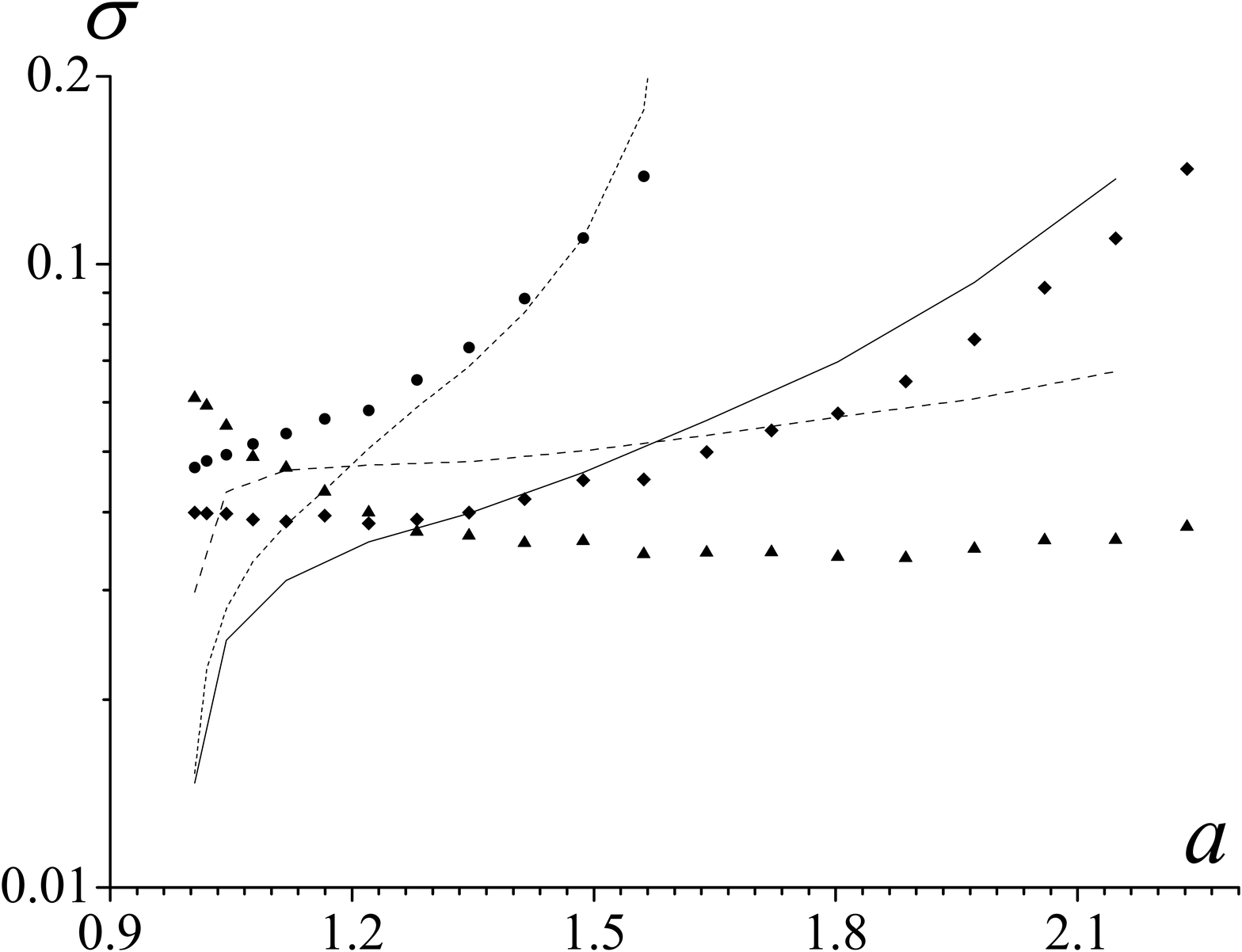}
\end{picture}
\parbox{230bp}{\caption{The SD versus current $a$ for different values of bias current
for $\gamma=0.001$. Circles, diamonds and triangles - computer simulations for
$i_0=0.5;0.7;0.9$, respectively.  Dashed, solid and long-dashed curves
- formula (\ref{sem}).}
\label{Fig6}}
\end{figure}

Very recently a universal, but approximate formula for the jitter $\sigma$ has been
derived in \cite{sem}. Since this formula has neither been compared with the results
of computer simulations nor investigated versus bias current of the clock generator we
have performed a certain analysis to test its validity for the considered task. In our
notations the formula for the jitter by Semenov and Inamdar \cite{sem} may be
presented in the form:
\begin{eqnarray}\label{sem}
\sigma=\sqrt{\frac{2\gamma}{\omega_c\tau}}\left|\frac{ d\tau}{ d i_0}\right|,
\end{eqnarray}
where $\tau$ is the MST, $i_0<1$ is the bias current of the junction and $\gamma$ is
the noise intensity.  The formula has been derived with the assumption that
fluctuations adiabatically follow the MST in working point that results in a jitter
of switching time. In spite of the simplicity of (\ref{sem}), it gives surprisingly
good coincidence with the results of computer simulations, see Fig.s 4 and 5,
especially for relatively large noise intensity $\gamma=0.1$ and smaller bias currents
0.5 and 0.7. For bias current $i_0=0.9$ and noise intensity $\gamma=0.001$ the error
may be up to $50-100 \%$ and the minimum of $\sigma$ versus $a$ observed in the
present paper is not reproduced.

We have considered fluctuational dynamics of a short JJ driven by a direct current and
SFQ pulse. It has been demonstrated that both mean switching time and its standard
deviation have minima as functions of the bias current of a clock generator,
responsible for the shape of SFQ pulse. Therefore, by proper choice of the bias
current of the junctions, both the response time and the jitter of the rapid
single flux quantum logic devices can be minimized.

Authors with to thank A.Yu. Kidiyarova-Shevchenko, K.K. Likharev and V.K. Semenov
for stimulating discussions and comments.

The work has been supported by the RFBR (Project No. 03-02-16533), INTAS (Project No.
01-0367), ISTC (Project No. 2445 and 3174) and by Russian Science Support Foundation.

\end{document}